# ON THE FRACTAL DIMENSION OF THE DUFFING ATTRACTOR

M. TARNOPOLSKI

Jagiellonian University, Astronomical Observatory, Orla 171, 30-244 Kraków, Poland
E-mail: mariusz.tarnopolski@uj.edu.pl

*Abstract*. The box counting dimension $d_C$ and the correlation dimension $d_G$ change with the number of numerically generated points forming the attractor. At a sufficiently large number of points the fractal dimension tends to a finite value. The obtained values are $d_C \approx 1.43$ and $d_G \approx 1.38$.

## 1. INTRODUCTION

The Duffing pendulum is a kind of a forced oscillator with damping and is governed by a nonlinear differential equation of the form

$$\ddot{x} + \delta \dot{x} + \beta x + \alpha x^3 = \gamma \cos(\omega_D t) \qquad (1)$$

where $\delta$, $\beta$, $\alpha$, $\gamma$, $\omega_D$ are constant parameters [1-3]. It is of great interest in nonlinear dynamics as it models many realistic, simple physical systems, e.g. a double spring pendulum, a spring pendulum with a nonlinear restoring force, i.e. not obeying the Hooke law, or a forced bar between two magnets [1, 4]. It can be constructed as various mechanical and electrical devices [5, 6]. Additionally, despite its apparent simplicity with only the $x^3$ nonlinear term, it exhibits a variety of behaviors. On the other hand, numerical integration can be easily conducted in order to examine the system in the means of the theory of chaos. Connected with the Duffing oscillator is the famous Duffing attractor, visible in a Poincaré surface of section presented in Fig. 1. The Poincaré section is a straightforward method for detecting chaos for given initial conditions of a dynamical system. In the case of forced systems the trajectory is sampled with a time step corresponding to the forcing frequency, e.g. $\omega_D$ in Eq. (1). If the solution is strictly periodic (allowing the presence of higher harmonics) it will appear as a finite number of points in the surface of section. A quasi-periodic oscillation will manifest itself *via* an analytical line due to the incommensurable frequencies. A chaotic solution will spread over a vast area of the phase space, forming a nontrivial set in the Poincaré surface of section. The Duffing attractor (being an invariant set, as it has the same structure for different initial conditions in the basin of attraction that lead to chaotic trajectories) is a fine example of a so called strange attractor, due to its fractal structure. This is because the system is dissipative, so a flow evolved droplet of initial conditions will asymptomatically converge to a zero Lebesgue measure set [7, 8]. Moreover, the Lyapunov spectrum [4, 7, 9-11] itself indicates the system to be dissipative, so the converged set is indeed a strange attractor. The Lyapunov Characteristic Exponents measure how fast nearby initial conditions diverge with time. If the divergence is fast enough, i.e. exponential, the system is called sensitive to initial conditions. This sensitivity is a necessary condition for the system to be chaotic. There are as many Lyapunov exponents as the dimension of the phase space. It is worthy transforming Eq. (1) into a autonomous dynamical system by setting two variables as the position and velocity $(x, \dot{x})$ respectively and the third one be the phase of the forcing, i.e. $\dot{\varphi} = \omega_D$ describes the evolution in the time-like direction of the phase space [1, 12]. This sets the dimension of the phase space equal to three, so there are three Lyapunov exponents. One of them, corresponding to the $\varphi$ direction, is equal to zero. Among the other two, one has to be positive for the system to be sensitive to initial conditions. So the last one has to be negative and have a greater modulus than the positive one in order for the system to posses a strange attractor. Also, a strange attractor is said to be self-similar, i.e. having a fractal structure (see Appendix A). It is a common procedure to verify the fractality of a set by estimating the Hausdorff (fractal) dimension [4, 13]. The fractal dimension gives information about how much of the space (e.g. the phase space or the space of stroboscopic variables used for constructing the Poincaré surface of section) is covered by the considered set. For instance, a set with a fractal dimension of 1.5 covers the space more densely than an analytical line but not as densely as a regular two-dimensional geometrical figure. Note that this does not mean that the higher the dimension, the larger area the fractal covers, like a circle is not *more two-dimensional* than a square, even if the latter one is smaller.

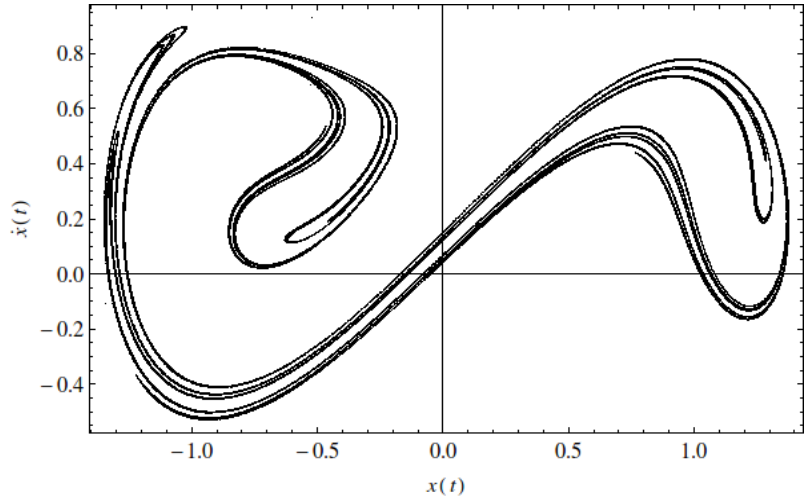

**Figure 1** - **The Duffing attractor formed of 286 478 points. Magnifications of this plot are presented in Appendix A.**

An ideal fractal should be formed of an infinite amount of points, although this is impossible to achieve in numerical computations. Also, the mathematical definition of the Hausdorff dimension does not provide a useful method for calculation. The most commonly used estimates are the box counting dimension and the correlation dimension [13]. It is obvious that if the examined set is formed of too few points, the fractal properties could not become apparent. So the set has to be consisted of a large enough number of points. Thus the fractal dimension depends on the amount of points.

This paper is organized in the following manner. Sec. 2 briefly presents numerical methods used for estimating the box counting and correlation dimensions. In Sec. 3 the results of estimating these dimensions for a numerically generated Duffing attractor are presented. In Sec. 4 concluding remarks are given.

## 2. METHOD

The fractal dimension is estimated in two ways: through the box counting and the correlation dimension. The box counting dimension is defined as

$$d_C = \lim_{\varepsilon \to 0} \frac{\ln N(\varepsilon)}{\ln(1/\varepsilon)} \quad (2)$$

where $N(\varepsilon)$ is the number of non-empty boxes (squares) of the size $\varepsilon$. This dimension was calculated using a computer algebra system Mathematica and the BoxCount function written by Pasquale Nardone [14].

The correlation dimension is defined as

$$d_G = \lim_{r \to 0} \frac{\ln C(r)}{\ln r} \quad (3)$$

with the estimate for the correlation function as

$$C(r) = \lim_{N \to \infty} \frac{1}{N^2} \sum_{i=1}^{N} \sum_{j=i+1}^{N} H(r - \|x_i - x_j\|) \quad (4)$$

where the Heaviside step function $H$ adds to $C(r)$ only points $x_i$ in a distance smaller than $r$ from $x_j$ and *vice versa*. The total number of points is denoted by $N$ here. In the following calculations $N$ is a finite value so the limit in Eq. (4) is omitted. The correlation sum is calculated using a parallel Python program.

Both limits in Eq. (2) and (3) are attained by using several small values of $\varepsilon$ and $r$ respectively and fitting a straight line to the obtained dependencies. The fractal dimension is estimated as the slope of the linear regression in both cases.

The numerical complexity of the correlation dimension algorithm is higher than the box counting one. Detailed comparative discussion is presented in Appendix C.

### 3. RESULTS

Eq. (1) was integrated using Mathematica's default method. Although the so called Clean Numerical Simulation (CNS) was recently developed [15-17 and references therein] in order to avoid truncation and round-off errors in a long time integration interval, herein obtained points forming the Duffing attractor are meant initially to form a set in stroboscopic variables. As is seen in Fig. 1 and 3, 4 in Appendix A, the attractor remains stable with the rise of $N$. Therefore it is claimed that for the purpose of this paper the CNS method is not necessary.

The following parameters were used:

$$\{\alpha,\beta,\delta,\gamma,\omega_D\}=\{1,-1,0.2,0.3,1\} \qquad (5)$$

Because β is negative, the potential is in a double-well form, so the system posseses two centers and one saddle as equlibrium points [3, 12]. The initial conditions were preselected to be $(x_0,\dot{x}_0)=(1,1)$. The attractor was formed by taking the values $(x,\dot{x})$ in stroboscopic variables with a step equal to $2\pi$ due to the forcing frequency $\omega_D=1$. The maximum length of the time series was equal to $1.8\cdot 10^6$ and lead to the attractor shown in Fig. 1.

Table 1
Fractal dimensions. The error in these quantities, estimated *via* the standard deviation of the linear regressions' slope, is not greater than 2%.

|  | Length of time series | Number of points forming the attractor | Box counting dimension | Correlation dimension |
|---|---|---|---|---|
| 1. | $1\cdot 10^3$ | 160 | 1.25164 | 1.45068 |
| 2. | $1\cdot 10^4$ | 1592 | 1.33008 | 1.35716 |
| 3. | $5\cdot 10^4$ | 7958 | 1.38010 | 1.39567 |
| 4. | $1\cdot 10^5$ | 15 916 | 1.39157 | 1.40250 |
| 5. | $2\cdot 10^5$ | 31 831 | 1.39461 | 1.38567 |
| 6. | $4\cdot 10^5$ | 63 662 | 1.41441 | 1.37372 |
| 7. | $6\cdot 10^5$ | 95 493 | 1.42120 | 1.37633 |
| 8. | $8\cdot 10^5$ | 127 324 | 1.41623 | 1.37722 |
| 9. | $1\cdot 10^6$ | 159 155 | 1.42120 | 1.37764 |
| 10. | $1.5\cdot 10^6$ | 238 733 | 1.42848 | 1.37895 |
| 11. | $1.8\cdot 10^6$ | 286 478 | 1.42515 | — |

Table 1 presents all lengths of obtained time series and the corresponding box counting and correlation dimension. Fig. 2 shows how do these fractal dimension estimates change with the number of points forming the attractor. The correlation dimension for $N$=286 478 was not calculated because the straightforward algorithm used is time and memory consuming.

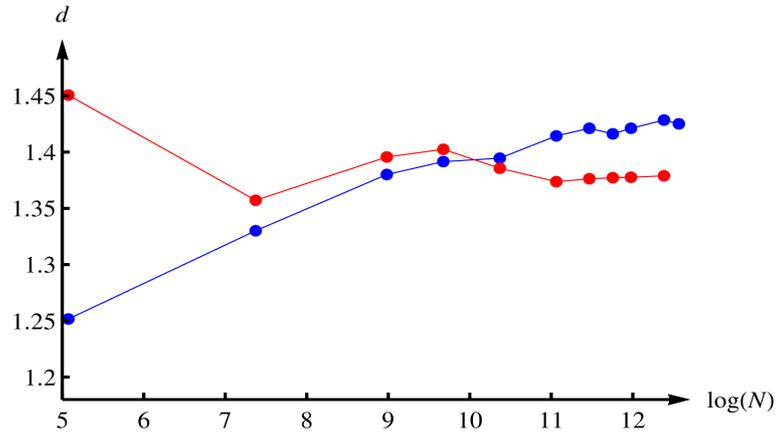

**Figure 2 - Fractal dimensions of the Duffing attractor. Blue marked points stand for the box counting dimension $d_C$ while the red marked ones symbolize the correlation dimension $d_G$. Note this is a semi-log plot.**

## 4. CONCLUSIONS

The fractal dimension was estimated for the Duffing attractor *via* the box counting and the correlation dimension. The values obtained are dependent on the number of points forming the examined set. In the box counting procedure one covers the space with boxes of a given size and counts how many of them are filled with points. This method relies only on global distribution of the points. On the other hand, the correlation dimension takes into account the local point density. Therefore, these two estimates can be expected to behave differently with the number of points varied.

The box counting dimension, in general, continually rose with the number $N$, although the bigger the $N$, the slower the rise was. It can be predicted that the $d_C$ value would finally reach its limit for $N$ large enough. The maximum $N$=286 478 appears to be a value large enough to give a reliable estimate of $d_C \approx 1.43$.

The rise of $d_C$ in Fig. 2 reaches its local maximum at $N$=95 493, followed by an oscillatory behavior, after which one can observe a decline at the maximum used value of $N$=286 478. This is due to the arbitrariness of the choice which part of the $\ln N(\varepsilon)$ vs. $\ln(1/\varepsilon)$ plot was linear (see Appendix B). Including or excluding one point from each fitting the monotonic behavior may be retained, although the linear regressions herein were performed so that the standard deviation of the slope was minimal. On the other hand, the relative difference in this case is less than 1%.

The correlation dimension even for an extremely small $N$=160 gave a value not greater than a few percent than the final one obtained for $N$=238 214, which was $d_G$≈1.38. What is a significant observation for numerical computations is that the $d_G$ for relatively small $N$ (starting herein from $N$=1592) does not differ much from its final value. After $N$=63 662 the relative changes in the $d_G$ value are not significant. Although the correlation dimension does not act monotonically on $N$ (as does not the box counting dimension), it appears to manifest some oscillating behavior for small $N$ and tends to a limit value for higher numbers of points. Therefore the final value of $d_G$≈1.38 is a good estimate for the fractal dimension.

The results show that the number of points forming the examined set plays a crucial role in reliably estimating the fractal dimension. Moreover, the correlation dimension acts more stable on the number of points $N$. Also, it takes into account the local distribution of points, so for a relatively small amount of points it gives a more reliable estimate for the fractal dimension.

*Acknowledgements.* The author is grateful to Slawomir Chrobak for the Python program that allowed to calculate the correlation dimension.

## 5. APPENDIX

### A. STRUCTURE OF THE DUFFING ATTRACTOR

Fig. 3 reveals the self-similar property of the Duffing attractor. This is a fundamental feature of all fractals and justifies the estimation of the fractal dimension in this paper.

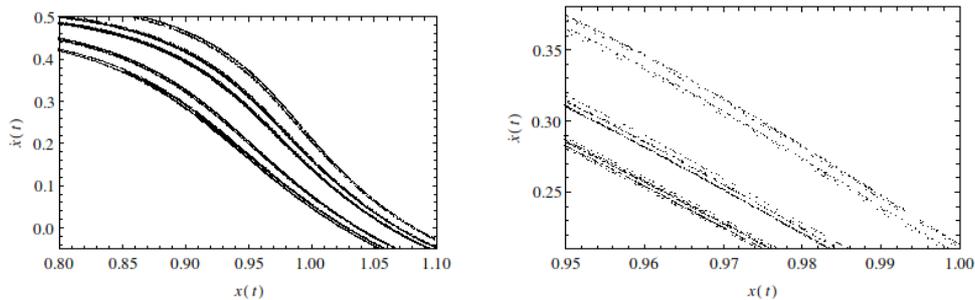

**Figure 3 - Magnifications of a part of Fig. 1 with a self-similar structure of the Duffing attractor.**

## B. LINEAR REGRESSIONS

An example of a result of a linear regression that lead to the box counting dimension for $N=1592$ is shown in Fig. 4a. The red line has a slope of $1.33008 \mp 0.02945$, while the blue one's slope is $1.28012 \mp 0.03560$, which was obtained by taking into account one point more than for the red line. This point is indicated by an arrow. The relative increase of the standard deviation is 21%, which is a significant value, although the relative decrease of the fractal dimension is only 4%. This means that within the error both values are equally reliable, however the criterion that was used to estimate the fractal dimension for each $N$ was so that the standard deviation was minimal.

All linear regressions, starting from $N=1592$, conducted in order to estimate the correlation dimension, had a form presented in Fig. 4b.

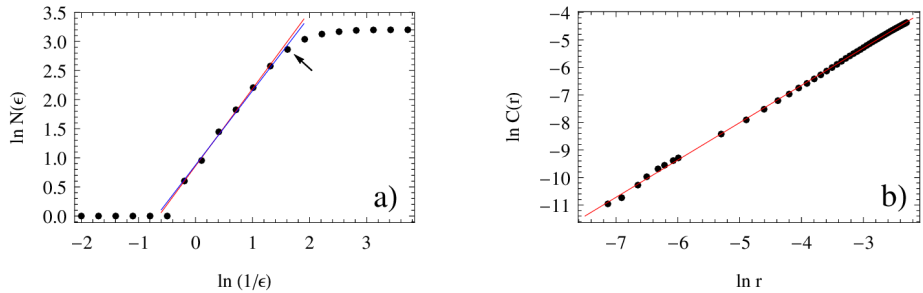

**Figure 4** - **Linear regressions for the attractor formed of $N=1592$ points that lead to the a) box counting dimension value of $d_C=1.33008$ and b) the correlation dimension $d_G=1.35716$.**

## C. NUMERICAL COMPLEXITY

The BoxCount function's evaluation time grows linearly with time, making the algorithm used an $O(n)$ one (Fig. 5a). This linearity arises from the fact that each loop's step runs over each point forming the attractor only once.

The correlation sum Python program has a complexity of $O(n^2)$ due to the fact that it calculates the distance beetwen all *pairs* of points, the number of which is exactly $n^2/2$. Using the fitted quadratic polynomial (Fig. 5b) the estimated time to conduct computations on 31 838 points is 12 943 s, while the actual time was 12 610 s, which is a fine correspondence.

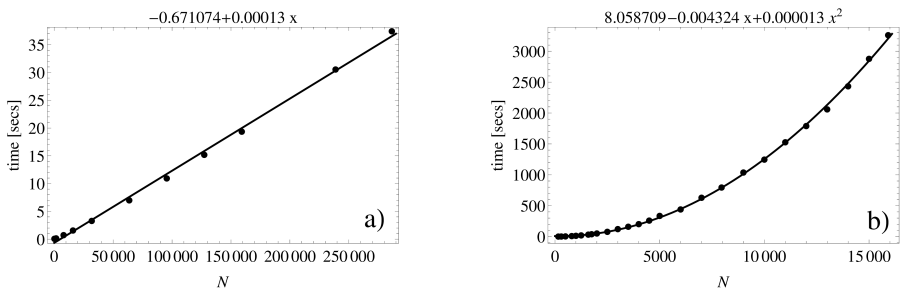

## D. THE CODE

The parallel Python program *corr.py* calculates the correlation sum from Eq. (4) neglecting the $1/N^2$ term. The input file *data.txt* should be a two-column list with the number of points in the first line. The *data* catalogue should be placed in the same directory as the *corr.py* file. *start*, *stop* and *step* are the minimal and maximal *r* values and the step between them, respectively. The output file *out.txt* is located in the *data* catalogue.

```python
#!/usr/bin/env python
# -*- coding: utf-8 -*-

from __future__ import print_function
import math
import multiprocessing
from datetime import datetime

CONFIG = {
    'input': 'data/data.txt',
    'start': 0.0025,
    'stop': 0.1,
    'step': 0.0025,
    'precision': 4
}
points = []
npoints = 0

def xfrange(start, stop, step, precision=2):
    r = start
    stop += step
    while r < stop:
        yield round(r, precision)
        r += step

def distance_between_points(xs, ys):
    xi, xj = xs
    yi, yj = ys
    x2 = (xi - xj) ** 2
    y2 = (yi - yj) ** 2
    d = math.sqrt(x2 + y2)
    return d

def heaviside_step(n):
    return int(n >= 0)

def _parse_lines(first, second):
    current_row = map(float, first.split())
    next_row = map(float, second.split())
```

```python
        xs, ys = zip(current_row, next_row)
        return xs, ys

    def calculate(input_queue, result_queue):
        while True:
            r = input_queue.get()
            h = 0
            try:
                for i in xrange(0, npoints):
                    for j in xrange(1 + i, npoints):
                        xs = points[i][0], points[j][0]
                        ys = points[i][1], points[j][1]
                        dbp = distance_between_points(xs, ys)
                        h += heaviside_step(r - dbp)
            except Exception, err:
                print(err)
            result_queue.put((r, h))
            input_queue.task_done()

    with open(CONFIG['input']) as f:
        try:
            for line in f:
                line = map(float, line.split())
                points.append(line)
            points.pop(0)
        except Exception, err:
            print(err)
        npoints = len(points)

    def run():
        steps = list(xfrange(CONFIG['start'], CONFIG['stop'], CONFIG['step'],
                    CONFIG['precision']))
        nworkers = multiprocessing.cpu_count()
        workers = []
        q = multiprocessing.JoinableQueue()
        rq = multiprocessing.Queue()

        for i in xrange(nworkers):
            p = multiprocessing.Process(target=calculate, args=(q, rq))
            p.daemon = True
            workers.append(p)

        for worker in workers:
            worker.start()

        for step in steps:
            q.put(step)
        q.join()
```

```
    results = []
    while not rq.empty():
       results.append(rq.get())
    results = sorted(results, key=lambda x: x[0])

    return results

def main():
    output = run()
    with open('data/out.txt', 'w') as f:
       f.write("r\tC(r)\n")
       for line in output:
          try:
             f.write("{0}\t{1}\n".format(line[0], line[1]))
          except:
             f.write("%.*f\t%d\n" % (CONFIG['precision'], line[0],
                              line[1]))

if __name__ == '__main__':
    start = datetime.now()
    main()
    print("Execution time {0}".format(datetime.now() - start))
```